\documentclass[aps,pre,11pt,twocolumn,reprint,longbibliography]{revtex4-1}

\usepackage{times}
\usepackage{graphicx}
\usepackage{amssymb}
\usepackage{amsmath}
\usepackage{bm}

\newcommand{\nn}{\nonumber}

\newcommand{\veps}{\varepsilon}

\makeatletter
\newcommand*{\balancecolsandclearpage}{%
  \close@column@grid
  \cleardoublepage
  \twocolumngrid
}
\makeatother

\begin{document}

\title{Experimental Demonstration of the Stabilization of Colloids by Addition 
of Salt}

\author{Sela Samin}
\author{Manuela Hod}
\author{Eitan Melamed}
\author{Moshe Gottlieb}
\author{Yoav Tsori$^*$}

\affiliation{Department of Chemical Engineering and the Ilse Katz Institute for 
Nanoscale
Science and Technology, Ben-Gurion University of the Negev, 84105 Beer-Sheva, 
Israel.\\
$^*$ To whom correspondence should be addressed; E-mail: tsori@bgu.ac.il}

\date{28 August 2014}

\begin{abstract}
We demonstrate a general non--Derjaguin-Landau-Verwey-Overbeek method to stabilize colloids in liquids. By 
this method, colloidal particles that initially form unstable suspension and 
sediment from the liquid are stabilized by the {\it addition} of salt to the 
suspending liquid. Yet, the salt is not expected to adsorb or directly interact 
with the surface of the colloids. For the method to work, the liquid should be a 
mixture, and the salt needs to be antagonistic such that each ion is 
preferentially solvated by a different component of the mixture. The 
stabilization may depend on the salt content, mixture composition, or distance 
from the mixture's coexistence line.
\end{abstract}
\maketitle 

\section{Introduction}

The stability of colloidal suspensions is important for the physical and 
chemical properties of pastes, paints and inks, and in a variety of other 
applications in material science. The van der Waals attraction between colloids 
can be overcome by steric repulsion, where surfactants or polymers are 
chemically or physically attached to the surface of the colloids and prevent 
them from aggregating \cite{pincus_mm1991}. In many cases these coatings are 
undesired because they change the surface chemistry, interfere with the activity 
of functional groups, block the contact between colloids once the suspension is 
dried, or affect the rheology of the liquid \cite{saville_book}. Alternatively, 
the colloidal dispersion may be stabilized electrostatically by means of charged 
molecules attached or adsorbed to the particle surface and the stability against 
aggregation is achieved by the Coulombic repulsion between the charged colloids. 
In the common Derjaguin-Landau-Verwey-Overbeek (DLVO) paradigm 
\cite{dlvo1,dlvo2}, when 
salt is added to an electrostatically-stabilized colloidal dispersion, the range 
of repulsion, set by the Debye length $\lambda_D$, decreases and the colloids 
tend to aggregate \cite{blaaderen_nature2003,blaaderen_nature2005}. 

In recent years, the importance of the preferential solvation of ions in 
different solvents has been realized \cite{sadakane2009,sadakane2013} and used in emulsions of two immiscible 
liquids. Leunissen {\it et al.} 
\cite{chaikin_pnas2007,van_roij_pccp2007,van_roij_prl2007} used the preferential 
solvation and resulting partitioning of antagonistic ion pairs to control the 
stability and organization of {\it charged} colloids. A recently published 
non-DLVO theory exploits ion solvation to predict that both neutral and 
charged colloids can be effectively suspended by the addition of 
salt \cite{sesc_jcp} to homogeneous mixtures. The key requirements are that (i) 
the suspending medium is a mixture of liquids and (ii) the salt is 
antagonistic; namely, the cation and anion are preferentially solvated in the 
different solvents \cite{marcus_book1}. The ions are not required to interact 
with the surface of the colloidal particle. The main advantage of the proposed 
stabilization method over the prevailing practice is that the 
stabilization 
is mediated by the liquid itself without the presence of large molecules or the 
need to modify the particle's surface. The theory predicts that the underlying 
mechanism should be effective in aqueous mixtures and it has a unique dependence 
on temperature, salt content, and mixture composition. Here we demonstrate 
experimentally, with a select set of experiments on key combinations of particles 
and liquids, that indeed colloidal stabilization can be achieved by the addition of 
antagonistic salts as predicted. 

In order to test the theory \cite{sesc_jcp}, we study experimentally the 
behavior of two types of neutral colloids: micron-sized cross-linked polystyrene 
(PS) microspheres and graphene sheets exfoliated from graphite particles. Two 
types of liquid mixtures are employed: a mixture of water and 2,6-lutidine 
which has a lower critical solution temperature \cite{grat93} and a mixture of water 
and acetonitrile which has an upper critical solution temperature (UCST) \cite{renard1965}. In the 
course of the experiments, we track the temporal behavior of the dispersions in 
each one of the two pure solution components alone, in the mixture in the 
presence of a nonantagonistic salt (NaCl), and in the presence of an 
antagonistic salt (NaBPh$_4$). The suspensions of the PS microspheres are 
studied by dynamic light scattering (DLS), whereas for the graphite or graphene 
suspensions visible-light transmission spectroscopy is used. Cryo-TEM imaging 
is carried out to complement the information regarding the aggregation state, 
Zeta-potential 
measurements to ascertain particle neutrality, and contact-angle measurements to 
determine the affinity of the mixture components to the solid surface.

According to the theory, the suspending efficiency of the proposed mechanism 
depends on the solvent-mixture composition, salt concentration, and the 
temperature $T$ in terms of its distance from the coexistence line temperature 
$T_t$. The sensitivity to the experimental variables is clearly demonstrated in 
Fig. 4 of Ref. \cite{sesc_jcp}. Rather than carry out an extensive search for the 
proper conditions, we use the theory, adjusted to correspond to the experimental 
systems at hand, to provide the guidelines for the judicious choice of the 
experimental parameters. In what follows, we provide a brief account of the 
theory \cite{sesc_jcp} with the required modifications and calculation 
pertaining to the experimental systems described above.

For colloids immersed in a generic mixture of two solvents, the total free energy 
is given as a sum of volume and surface contributions $F=\int\left(f_m+f_{\rm 
es}+f_{\rm ion}\right)d{\bf r}+\int f_\gamma d{\bf 
r}_s$, where ${\bf r}_s$ is a vector on the surface. $f_m$ is the mixing free 
energy density given by \cite{safran_book}
\begin{align} 
a^3f_m&=k_BT\bigl[\phi\log\phi+(1-\phi)\log(1-\phi) \nn \\ 
&+\chi\phi(1-\phi)\bigr]+\frac12
C(\boldsymbol{\nabla}\phi)^2~.
\end{align}
Here $\phi$ is the {\it local} mole fraction of the more polar solvent, $k_B$ is 
the Boltzmann constant, $T$ is the temperature, $\chi\sim 1/T$ is the 
Flory-Huggins interaction parameter, $a$ is a molecular length, and $C$ is a 
constant.
The short-range interfacial interaction of the mixture with the solid colloid 
surface is given by $f_\gamma=\Delta\gamma\phi({\bf r}_s)+\sigma\psi({\bf 
r}_s)$, where $\Delta\gamma$ is the difference between the surface tensions of 
the two liquids and the solid (assumed chemically homogeneous) and $\sigma$ is 
the surface charge density of the colloid. A positive $\Delta\gamma$ means the surface prefers 
the less polar cosolvent. The new stabilization mechanism is not based on 
specific interactions between the ions and the surface 
\cite{hunter2001,onuki2011}, and hence these are not included here. The 
electrostatic energy density $f_{\rm es}=-(1/2)\veps(\phi) (\boldsymbol{\nabla}\psi)^2$ is 
expressed by the local electrostatic potential $\psi$ and the constitutive 
relation between the dielectric constant and mixture composition $\veps(\phi)$, assumed linear. 

The Gibbs transfer energy of moving an ion from one solvent to another gives 
rise to numerous important interfacial phenomena 
\cite{levin_prl2009a,levin_prl2009b,nellen2011} and can even lead to 
flocculation of colloids \cite{netz_prl1996}. In the present context and for a 
monovalent salt, the ions' entropy, electrostatic energy, and preferential 
solvation is modeled by
\begin{align} 
f_{\rm
ion}&=k_BT\left\{n^+\left[\log(a^3n^+)-1\right]+n^-\left[\log(a^3n^-)-1\right]\right\} \nn \\
&+e(n^+-n^-)\psi-k_BT\left(\Delta u^+n^++\Delta u^-n^-\right)\phi~.
\end{align}
Here $n^\pm$ are the average ion number density in the system, and $e$ is the elementary charge. The 
parameters $\Delta u^\pm$ express the affinity of the positive and negative ions 
toward the polar phase \cite{efips_epl,onuki2011,bier2011,van_roij_pccp2007}. When the 
antagonistic salt used here, NaBPh$_4$, is added to the mixture of water and 
2,6-lutidine the Na$^+$ cation is hydrophilic with $\Delta u^+\simeq 6$, while the 
BPh$_4^-$ anion is hydrophobic with $\Delta u^-\simeq -16$ 
\cite{persson1990,comment}, and thus both requirements (i) and (ii) listed above hold. The colloids will be 
stabilized irrespective of the sign of $\Delta \gamma$ or $\Delta u^\pm$ and as 
long as $|\Delta u^+-\Delta u^-|$ is large enough.

In equilibrium the composition $\phi$, ion densities $n^\pm$, and electrostatic 
potential $\psi$ satisfy the three coupled equations $\delta F/\delta\phi=0$, 
$\delta F/\delta n^\pm=0$, and the Poisson equation $\delta F/\delta\psi=0$. The 
boundary conditions at the colloid surface are ${\bf n}\cdot\boldsymbol{\nabla}\psi=\sigma/\veps(\phi)$ and ${\bf 
n}\cdot\boldsymbol{\nabla}\phi=-\Delta\gamma/C$, where ${\bf n}$ is a unit vector normal to 
the colloid surface.
We calculate the interaction between two neutral spherical colloids of radius 
$R$ separated by a distance $D$ and immersed in a mixture at average composition 
$\phi_0$ and ion density $n_0$ as follows. We begin by solving the governing 
equations for two flat plates a distance $D$ apart. Once the equilibrium 
profiles are known the pressure tensor
$P_{ik}=\left(\phi \delta f/\delta\phi+n^+\delta f/\delta
n^++n^-\delta f/\delta n^--f\right)\delta_{ik}-\veps E_iE_k$
 is obtained \cite{landau2,andelman_jpcb2009}. Here ${\bf E}=-\boldsymbol{\nabla}\psi$ is the electric field.
We define $\Omega$ as the integral of the osmotic pressure from $D$ to $\infty$. 
When the distance between the colloids is much smaller than their size, $D\ll 
R$, Derjaguin's approximation holds and the total effective colloid potential is 
\cite{saville_book}
\begin{equation} \label{eq_U_of_D}
U(D)=\pi R\int_D^\infty\Omega(D')dD'-\frac{AR}{12D}~,
\end{equation}
where $A$ is Hamaker's constant. In Eq. \ref{eq_U_of_D}, we use a simple form of 
van der Waals interaction, not taking into account the wetting close to the 
colloid \cite{Beysens1998} and screening by the salt \cite{parsegian_book}. The 
wetting layer around a hydrophobic colloid in the water--2,6-lutidine mixture 
increases the van der Waals attraction \cite{Beysens1998}, though in our 
experiments the effect is expected to be small, since the temperature is kept 
relatively far from $T_c$. In addition, as will be demonstrated below, we do 
not observe stable suspension when the hydrophilic salt (NaCl) is added, and 
this result implies that salt screening of van der Waals interaction alone is not 
enough to stabilize the colloids. We stress that the aim of the simple theory we 
use is to isolate one possible mechanism and show that it can be comparable to or 
even larger than others in the stabilization.

Insight can be gained by a standard linear theory valid when $e\psi\ll k_BT$ and 
$|\Delta u^\pm(\phi-\phi_0)|\ll 1$ hold. The potential and composition are then 
given as a sum of four exponentials $e^{\pm q_iz}$ with the two wave numbers 
$q_i$ ($i=1,2$) given by $q_1^2q_2^2=1/(\xi\lambda_D)^2$ and
$q_1^2+q_2^2=1/\lambda_D^2-1/\xi^2-2n_0\left(\Delta u^+-\Delta u^-\right)^2/C$, 
where $\xi$ is the correlation length modified by the salt 
\cite{onuki2011,nellen2011,bier2012}. For electrically neutral colloids ($\sigma=0$), the 
amplitudes of the exponentials are proportional to the difference in wettability 
of the two solvents, $\Delta\gamma$, and the height of the barrier $U(D_{\rm 
max})$ is proportional to $(\Delta\gamma)^2$. When a value of approximatly $10$ m$M$ of 
salt is used in the analytical expressions, it follows that the barrier location 
$D_{\rm max}$ is in the range $5$--$15$ nm and the barrier height can be 
significantly larger than approximatly $3 k_BT$ thus preventing colloidal coagulation.

\section{Experimental Details}

\subsection{Preparation and characterization of polymer colloids}

The spherical colloidal particles are synthesized by using distillation-precipitation 
polymerization of divinylbenzene \cite{bai2007}. The size distribution and shape 
are determined by means of scanning electron microscopy using JSM-7400F (JEOL) 
ultrahigh-resolution cold field-emission gun SEM; see Fig. S2 in the Supplemental Material 
\cite{supplementary}.

For the surface charge measurement, the colloids are dispersed in ethanol, and 
their $\zeta$ potential is measured on a Zetasizer Nano ZS (Malvern) at $298$ 
K using a universal dip cell. The measured peak value of the $\zeta$ potential is 
$0\pm2$ mV.

\vspace{0.5cm}
\subsection{DLS of polymer particles}

The mixture of water (deionized to a resistivity of $18.2$ M$\Omega$ cm) and 
2,6-lutidine (Sigma-Adrich, purified by redistillation, $\geq99\%$) with $71$ wt\% 
water is prepared at ambient temperature. The addition of the polymer 
microspheres to the samples for the DLS experiments is performed in two steps. 
First, the colloids are weighed and $1$ ml of the water--2,6-lutidine mixture 
and salt are added. The sample containing the particles is then placed in the 
ultrasonic bath and sonicated for $15$ min. After sonication, it is diluted with 
the same water-lutidine mixture, shaken, and further sonicated for $3-5$ min. 
Subsequently, it is equilibrated at the required temperature ($T=T_t-6$ K) by 
waiting for $10$ min. 

DLS is measured on the CGS-3 equipped with a LSE-5004 cross-correlator (ALV, 
Germany) at a constant angle of $90^\circ$ using clear glass vials 
(Sigma-Aldrich). Phase diagrams are separately determined for the mixture in 
the absence of salt and after the addition of $20$ m$M$ NaBPh$_4$ (see Fig. S5 in the 
Supplemental Material \cite{supplementary}). 
For the solution mixture in the absence of salt at a water weight fraction of 0.71 
($\phi$=$0.935$), $T_t=307$ K, and experiments are carried out at $T=301$ K. With 
the salt $T_t=321$ K, and experiments are carried out at $T=315$ K. During the 
experiment, the temperature is kept constant to within $0.1$ K by using a Julabo 
CF31 thermostat. Each measurement consists of five runs of $10$e each and 
is carried out at a beam wavelength of $632.8$ nm. The CONTIN algorithm is 
used for the extraction of the data. The Stokes-Einstein relation is employed 
for the calculation of the hydrodynamic radius $R_h$. Since temperature and 
salt impact the viscosity, we determine the viscosity for each sample by using a 
Cannon-Fenske viscometer at the appropriate temperature and composition: (i) no 
salt mixture -- $T=301$ K, $\eta$=$2.25$ mPa s; (ii) mixture with $20$ m$M$ 
NaBPh$_4$ -- $T=315$ K, $\eta$=$1.98$ mPa s; (iii) mixture at $T=305$ K, 20 m$M$ antagonistic 
salt -- $\eta$=$2.12$ mPa s; 25 m$M$ salt -- $\eta$=$2.16$ mPa s. 
The mass-weighted DLS data are smoothed using cubic interpolation. The synthesis 
yields also smaller particles with diameter $<500$ nm. The contribution of 
these small colloids to the total measured intensity is negligible and therefore not 
shown.

\vspace{0.5cm}
\subsection{Preparation of graphene}

Graphite flakes are purchased from 
Sigma-Aldrich and used as received. For each experiment $0.1$ wt\% are mixed with 
the respective mixture and placed in a sonicator bath (Elma, S $10$ Elmasonic, 
Germany) for 3 h to allow exfoliation. The sonicator is operated under 
ice-cooling in order to prevent heating and keep the temperature at approximatly $273$ K 
throughout the process. To separate the resulting graphene sheets from 
nonexfoliated graphite flakes, samples are centrifuged at a constant temperature 
(Hermle Z$383$K, R-max $9.6$ cm, Germany, $277$ K) at a rate of $1000$ rpm for $5$ 
min. 

\vspace{0.5cm}
\subsection{Visible-light transmission in graphene dispersions}

The visible-light transmission through the graphene dispersions is measured by 
a Jasco V-$530$ UV-visible spectrometer at a wavelength of $660$ nm. We measure the 
transmission in different mixtures of water (deionized to a resistivity of $18.2$ 
M$\Omega$ cm) and acetonitrile (Sigma-Aldrich, anhydrous $99.8\%$). Transmission 
values are converted into graphene concentration by using an extinction 
coefficient of $3.0\times10^3$ l g$^{-1}$ m$^{-1}$ taken from the literature 
\cite{regev2013}. The error in the concentration is estimated from the standard 
deviation of several measurements for a mixture with $20$ wt\% acetonitrile and 
the antagonistic salt.

\section{Results}

\begin{figure*}[t!]
\begin{center}
\includegraphics[clip,width=0.8\textwidth]{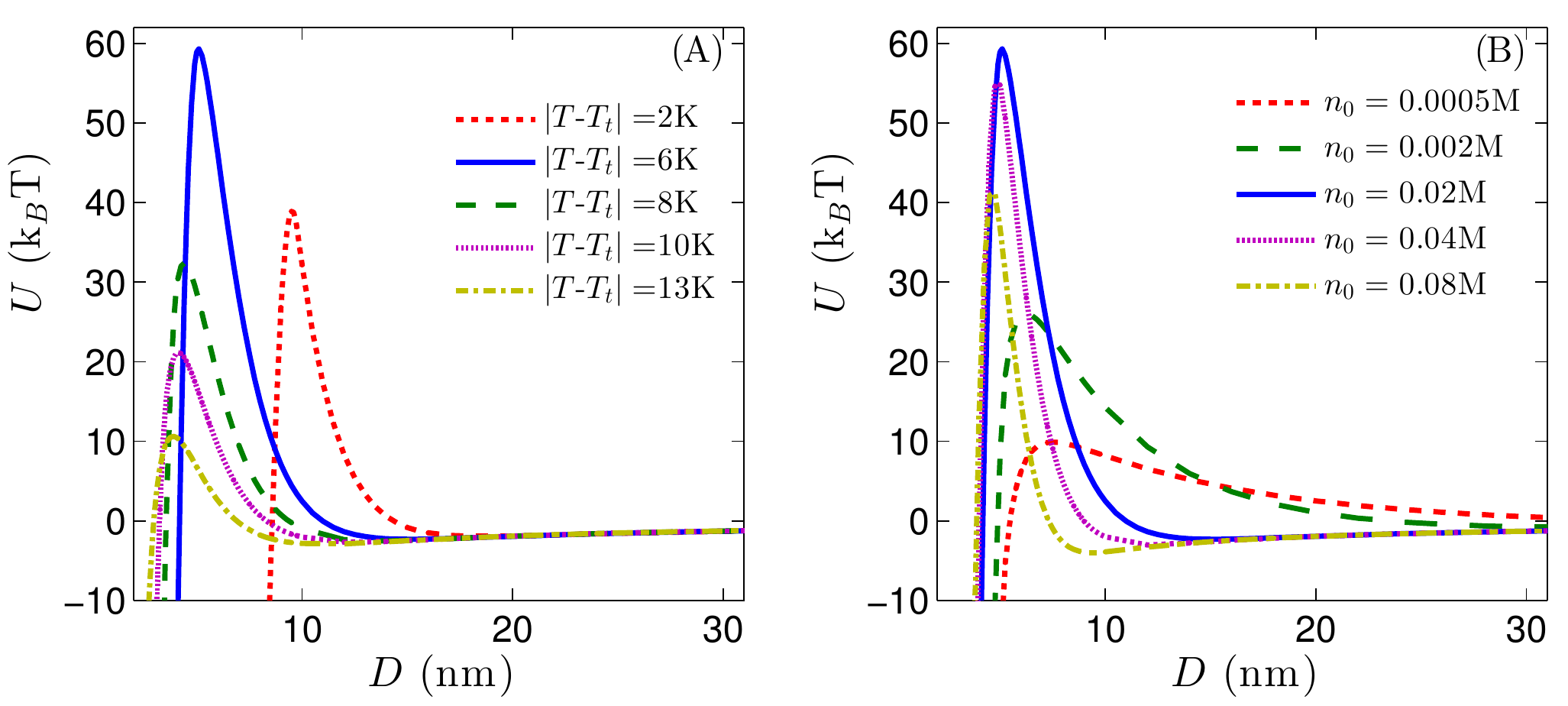}
\caption{\small Calculated effective potential between two colloids $U(D)$ from 
Eq. (\ref{eq_U_of_D}) vs colloid surface separation $D$ for varying 
temperatures $T$ (a) and salt concentrations $n_0$ (b). The location of the 
barrier peak at $D_{\rm max}$ decreases with increasing distance from the 
coexistence line $|T-T_t|$ or with increasing salt $n_0$ ($\lambda_D$ 
decreases). The mixture composition is such that $|T_c-T_t|=10$ K. In (a) 
$n_0=20$ m$M$ is constant while in (b)  $|T-T_t|=6$ K is constant. For the 
water--2,6-lutidine mixture containing the antagonistic salt NaBPh$_4$ we used 
$T_c=311$ K, $a=3.4$ \AA{}, $C=\chi/a$, $\veps_{\rm 2,6-lutidine}=6.9$, 
$\veps_{\rm water}=79.5$, $\Delta\gamma=0.1k_BT a^{-2}$, and $\Delta 
u^+=-\Delta u^-=8$. For the colloidal PS particles we used $R=1$ $\mu$m and 
$A=2\times10^{-21}$ J.
}
\label{fig1}
\end{center}
\end{figure*}

Because of the coupling between the mixture composition, salt concentration, and 
temperature (distance from coexistence line $|T-T_t|$) we carry out a somewhat 
crude iterative process. The full nonlinear potential is calculated repeatedly 
for different salt concentrations and temperatures. Eventually, we settle on a 
set of values which, conceded, are not optimized but provide the essence of the 
stabilization mechanism. Figure \ref{fig1} shows the full nonlinear potential 
$U(D)$ vs temperature and salt content. The solid surface has a short-range 
chemical attraction to one of the solvents. For a hydrophobic colloid, water is 
depleted from its surface, and due to the preferential solvation of the ions, the 
vicinity of the surface has more hydrophobic than hydrophilic ions [see Figure 
S1 (a) in the Supplemental Material \cite{supplementary}]. In the case of a hydrophilic colloid, the more hydrophilic ions are 
enriched near it. Thus, the colloid becomes effectively charged when {\it either} 
of the solvents is adsorbed at the surface and entropically driven repulsion 
appears at large distances 
\cite{chaikin_pnas2007,van_roij_prl2007,van_roij_pccp2007}. The potential has a 
barrier at $D=D_{\rm max}$ and is attractive when $D<D_{\rm max}$ due to van 
der Waals attraction and critical adsorption
\cite{evans1986}. As salt is added, $D_{\rm max}$ shifts to lower values while 
the potential barrier increases. Guided by the results of the calculation we 
opt to use a salt concentration of $n_0\approx 20$ m$M$ and $|T-T_t|\approx 6$ 
K. At this point, to further refine the correspondence between theory and 
experiment, the phase diagram for water--2,6-lutidine with 20 m$M$ of NaBPh$_4$ is 
obtained ($\phi_c=0.879$ and $T_c=311$ K; see Fig. S5 in the Supplemental Material \cite{supplementary}). Relying on 
the phase diagram and the fact that the colloids are hydrophobic, we select 
for the experiments a solvent mixture with a water weight fraction of 0.71 
($\phi_0=0.935$), for which $T_t=321$ K and hence $|T_c-T_t|=10$ K as in Fig. 
\ref{fig1}. The calculated salt-concentration dependence at the selected 
temperature and mixture composition is shown in Fig. \ref{fig1}(b), pointing 
towards the choice of 20 m$M$ salt concentration.

\subsection{Experimental validation I} 

Cross-linked polystyrene colloids are prepared by using distillation-precipitation 
polymerization. Their radius is determined by scanning electron microscopy to 
be $R=0.85\pm 0.3$ $\mu$m, and their $\zeta$ potential is found to be zero. The 
colloids and antagonistic (NaBPh$_4$) or nonantagonistic (NaCl) salts are 
added to the mixture of water--2,6-lutidine.

\begin{figure*}[!th]
\begin{center}
\includegraphics[width=0.68\textwidth]{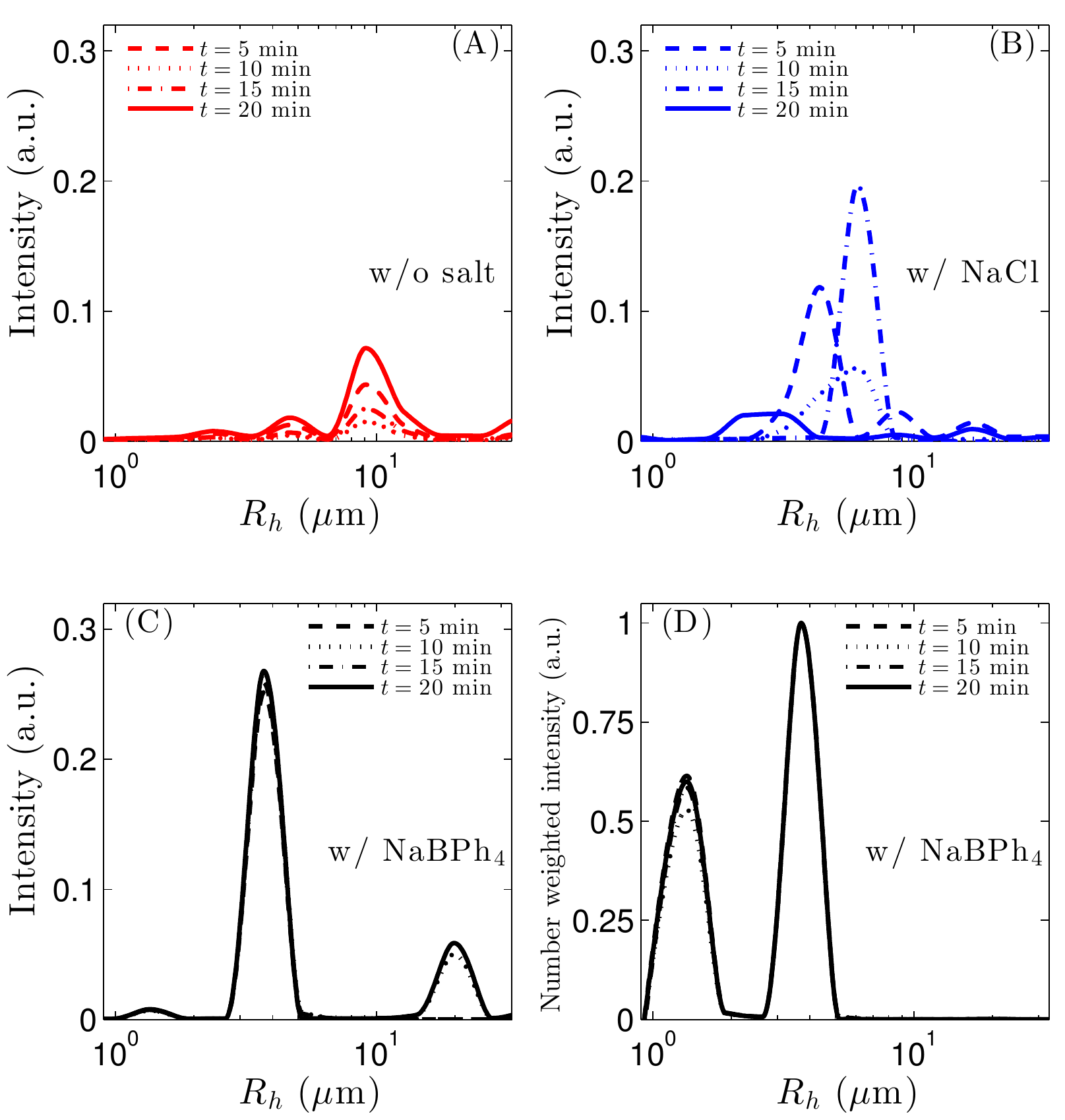}
\caption{\small (a)-(c) Colloidal mass-weighted distribution vs hydrodynamic 
radius $R_h$, as obtained by dynamic light scattering. Curves show 
distributions of polystyrene colloids in mixtures of water and 2,6-lutidine at 
different times (a) when no salt is added, (b) with $20$ m$M$ of NaCl, and (c) with 
$20$ m$M$ of the antagonistic salt NaBPh$_4$. Without salt or with NaCl, colloids 
form aggregates that grow in time and eventually sediment from the solution. 
With the antagonistic salt, the aggregates are small and stable for the duration 
of the experiment. (d) Number-weighted distributions for the suspension with 
NaBPh$_4$. All measurements are at $|T-T_t|=6$ K.
}
\label{fig2}
\end{center}
\end{figure*}

Visual inspection shows that the colloids coagulate and sediment in pure 
lutidine, in the mixture without salt, or when the salt is not antagonistic 
(NaCl). However, when the antagonistic salt is added, the colloids remain 
suspended over a long time. 
When the antagonistic salt is added to pure lutidine, the colloids 
coagulate immediately. This result supports the notion that there is no significant direct 
chemical interaction of the ions with the colloids. These qualitative 
observations are quantified by DLS for samples at 
$|T-T_t|=6$ K. Fig. \ref{fig2}(a) shows that the distribution of sizes shifts 
in time to larger aggregates when no salt is added to the mixture. The 
distributions in Fig. \ref{fig2}(b), corresponding to an addition of $20$ m$M$ 
NaCl, are similar in nature, and the size of aggregates shifts to larger values 
with increasing time until they sediment. This behavior is commonplace for 
neutral or slightly charged particles, and it is revealed in our numerical 
calculation \cite{sesc_jcp} assuming hydrophilic salt ($\Delta u^+=\Delta u^-$) 
even if the colloids' surface potential is as large as $30$ mV. A very different 
behavior is observed for the antagonistic 
salt, as shown in Fig. \ref{fig2}(c) and Fig. \ref{fig2}(d), where the aggregates are small and 
stable throughout the entire measurement period of $20$ min.

\begin{figure*}[!th]
\begin{center}
\includegraphics[width=0.682\textwidth]{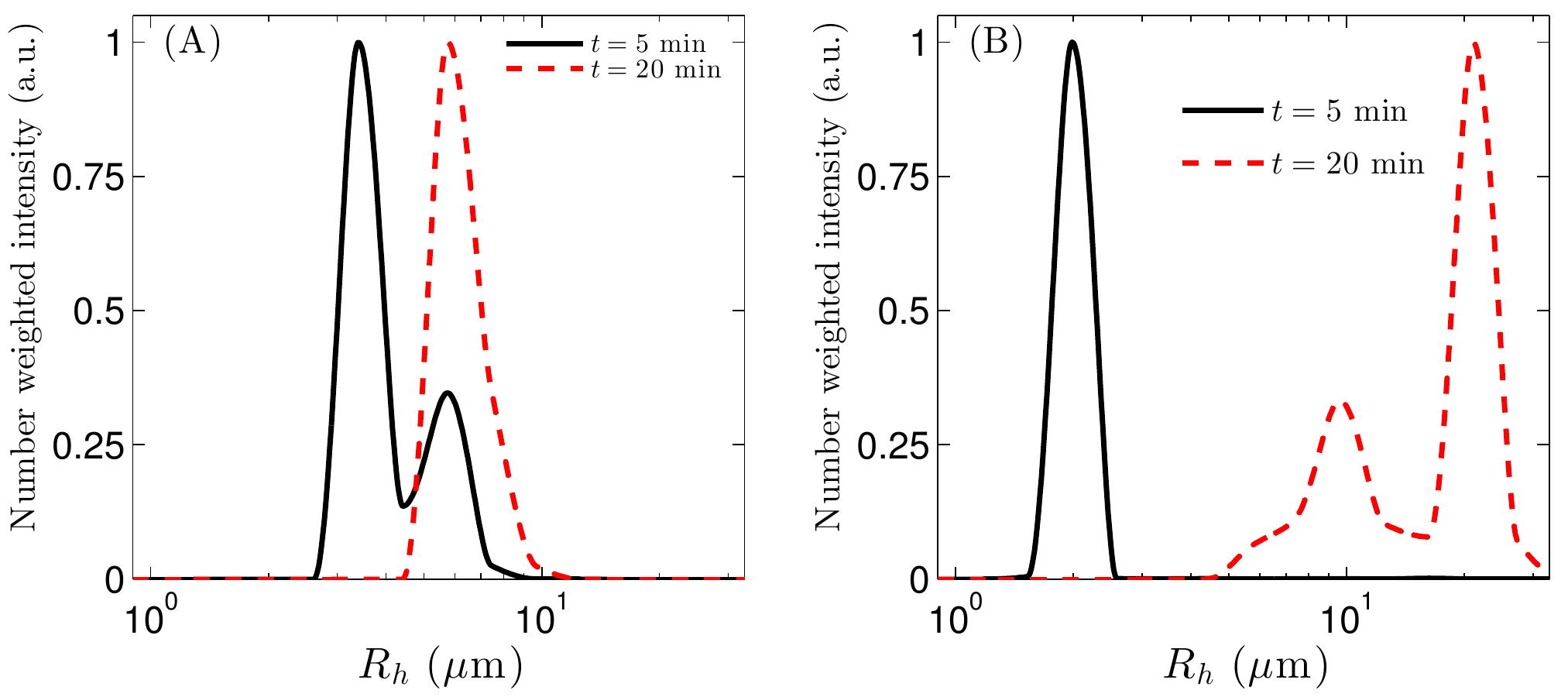}
\caption{\small Colloidal number-weighted distribution vs hydrodynamic radius 
$R_h$, as obtained by DLS. Curves show distributions of 
polystyrene colloids in mixtures of water and 2,6-lutidine at $T=305$ K. (a) With $20$ m$M$ of the antagonistic salt NaBPh$_4$, $|T-T_t|=16$ K, and (b) with $25$ m$M$ of 
the antagonistic salt. In both cases, colloids form aggregates that grow in 
time and eventually sediment from the solution.
}
\label{fig3}
\end{center}
\end{figure*}

Based on the theoretical calculations shown in Fig. \ref{fig1}(a) the energy 
barrier for aggregation at $|T-T_t|>13$ K should be very low, favoring 
aggregation. This result is confirmed by examining a 20-m$M$ and 25-m$M$ solutions of the 
antagonistic salt at $T=305$ K ($|T-T_t|=16$ K) as clearly indicated by the 
increase in size of the aggregates depicted in Fig. \ref{fig3}.

\subsection{Experimental validation II}

To demonstrate the wide scope of the dispersion principle, we test it with a 
different type (carbonaceous) and shape (sheets) of colloid and different 
solvent mixture (UCST type). Measurements are carried out at room temperature 
in a mixture of water and acetonitrile (UCST, $T_c\simeq 272$ K at a water mole 
fraction $\phi_c\simeq 0.64$). Carbon-based colloids, which disperse in pure 
2,6-lutidine, do not disperse in pure water or in acetonitrile.

\begin{figure}[t!]
\begin{center}
\includegraphics[clip,width=0.4\textwidth]{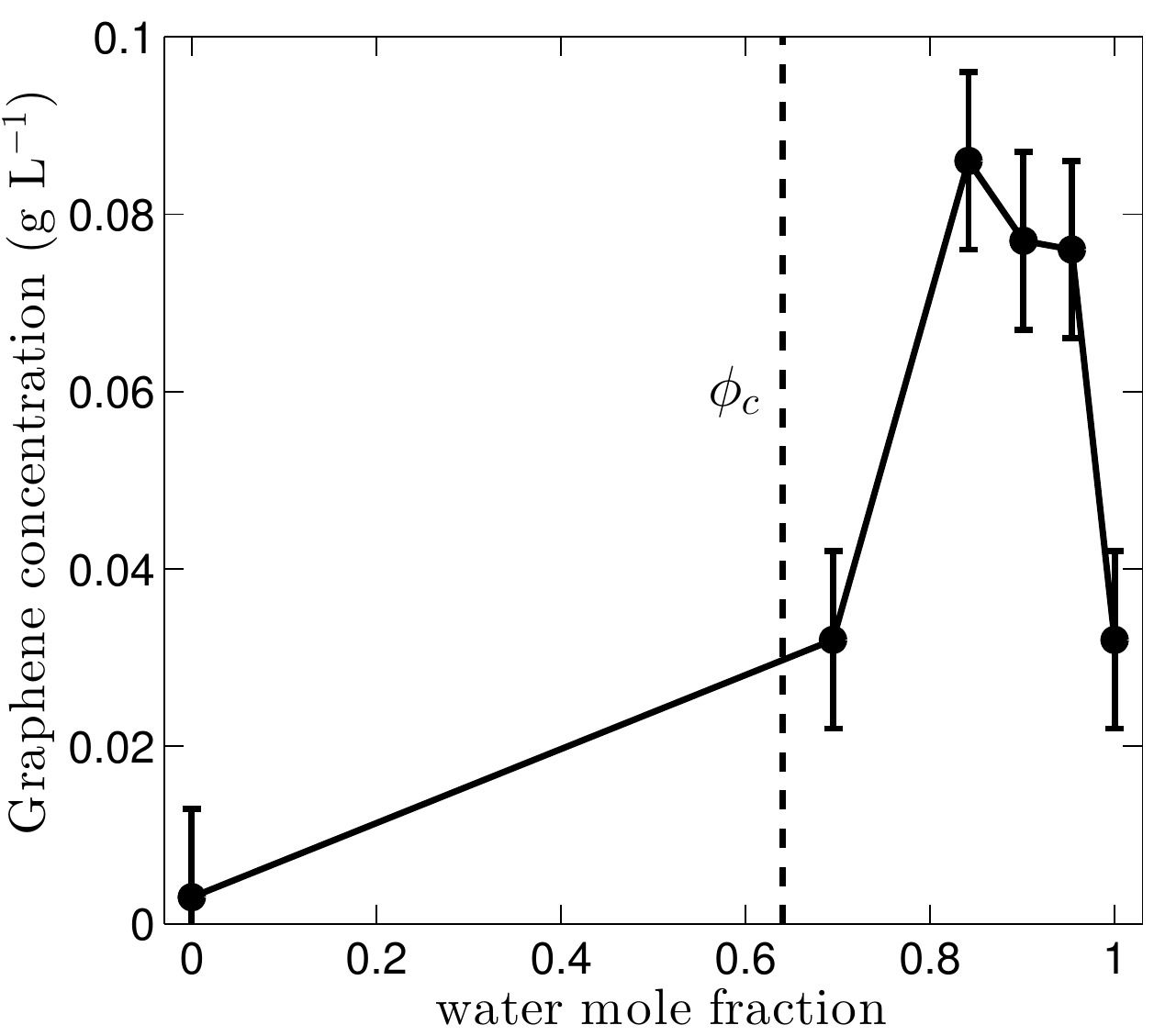}
\caption{\small Concentration of graphene dispersed in mixtures of water and 
acetonitrile containing $20$ m$M$ of NaBPh$_4$ at $t=0.5$ h. UV-visible transmission measurements are performed at room temperature $|T-T_t|>25$ K. The dashed line is the critical water mole fraction. As clearly observed at concentrations above critical, $0.85\lesssim\phi\lesssim0.95$, exfoliated graphene sheets remain dispersed in the liquid mixture. Dispersed graphene sheets are found even after 3 months.}
\label{fig4}
\end{center}
\end{figure}

We use ultrasonication to exfoliate graphene from graphite flakes. UV-visible 
spectrometry of the suspensions is performed by measuring the transmission 
intensity at a wavelength of $660$ nm. The transmission values are converted to 
concentration estimates with the Beer-Lambert law by using a literature value for 
the extinction coefficient \cite{regev2013}. For a mixture ($80$ wt\%, 
$\phi=0.90$) with $20$ m$M$ NaCl or in the absence of salt, the transmission 
after $t=0.5$ h is $100\%$ within the experimental error and therefore no 
dispersion of graphene is obtained. The same holds for pure acetonitrile and 
water. However, as shown in Fig. \ref{fig4}, the addition of an antagonistic 
salt results in the exfoliation of graphite, and a dispersion of graphene is 
obtained at mixture compositions $\phi \gtrsim \phi_c$ (see also 
Fig. S4 in the Supplemental Material \cite{supplementary}). A small amount of graphene ($\approx 0.03$ g l$^{-1}$) is dispersed 
even in pure water with NaBPh$_4$. We speculate that this dispersion is due to a specific 
interaction of the 
hydrophobic BPh$_4^{-}$ ions and the graphite, which can be incorporated into the 
theory \cite{onuki2011}. The dispersion is enhanced more than twofold when the 
antagonistic salt is added to mixtures in the composition range 
$0.85\lesssim\phi\lesssim0.95$. This increase cannot be accounted for by an 
ion-surface interaction, since BPh$_4^{-}$ ions are expected to adsorb less in the 
mixture containing a nonpolar component.
No stabilization is observed when the antagonistic salt is added to pure 
acetonitrile. The optimal dispersion at off-critical and water-rich mixtures is 
in agreement with the theoretical prediction in Fig. 4 of Ref. \cite{sesc_jcp} 
for a hydrophobic surface. The concentration of graphene decays slowly to about 
$25\%$ of its value at $t=0.5$ h within $24$ h. However, for the 
sample with $\phi=0.901$, many graphene sheets are clearly observed in cryo-TEM 
images taken 3 months after the sample preparation (see Fig. S4 in the Supplemental Material \cite{supplementary}), 
indicating a stable dispersion.

\balancecolsandclearpage

\subsection{Experimental validation III} 

When NaBPh$_4$ is added to the mixture 
of water, acetonitrile, and 
cross-linked-polymer microspheres, the concentration of dispersed polymer colloids 
is increased by a mere $20\%$--$30\%$. The reason for this relatively modest 
increase compared to the one observed in a mixture of water and 2,6-lutidine can 
be traced to the fact that acetonitrile is significantly less hydrophobic than 
lutidine, and hence the value of $\Delta\gamma$ is too small for an effective 
stabilization; see Fig. S3 in the Supplemental Material \cite{supplementary}. 

Dispersion of colloids by the addition of salts is a versatile method and 
straightforward in practice. It has a unique dependence on temperature and 
composition and does not rely on direct adsorption of the ions on the colloids. 
We do not attempt to optimize the preparation protocol (e.g., longer or repeated 
tip sonication, different salt concentrations or variations in temperature or in 
mixture composition). Such an optimization is expected to greatly enhance the 
dispersion efficiency, and thus the method outlined above could be potentially 
useful in many cases where surfactants or grafting with polymers is inadequate. 
For example, graphene is conventionally dispersed with surfactants and is 
subsequently spin-coated on a substrate. Currently, the coating is transparent, 
but the surfactants degrade the in-plane conductivity; we speculate that 
replacing the surfactants with salt that is not adsorbed to the particle's 
surface will increase this conductivity significantly.

\section*{Acknowledgments}

Y. T. acknowledges support from the European 
Research Council
``Starting Grant'' No. 259205, COST Action MP1106, and Israel Science Foundation 
Grants No. 11/10 and No. 56/14.

S.S. and M. H. contributed equally to this work.

% \bibliography{refdb_xiv}

\begin{thebibliography}{35}%
\makeatletter
\providecommand \@ifxundefined [1]{%
 \@ifx{#1\undefined}
}%
\providecommand \@ifnum [1]{%
 \ifnum #1\expandafter \@firstoftwo
 \else \expandafter \@secondoftwo
 \fi
}%
\providecommand \@ifx [1]{%
 \ifx #1\expandafter \@firstoftwo
 \else \expandafter \@secondoftwo
 \fi
}%
\providecommand \natexlab [1]{#1}%
\providecommand \enquote  [1]{``#1''}%
\providecommand \bibnamefont  [1]{#1}%
\providecommand \bibfnamefont [1]{#1}%
\providecommand \citenamefont [1]{#1}%
\providecommand \href@noop [0]{\@secondoftwo}%
\providecommand \href [0]{\begingroup \@sanitize@url \@href}%
\providecommand \@href[1]{\@@startlink{#1}\@@href}%
\providecommand \@@href[1]{\endgroup#1\@@endlink}%
\providecommand \@sanitize@url [0]{\catcode `\\12\catcode `\$12\catcode
  `\&12\catcode `\#12\catcode `\^12\catcode `\_12\catcode `\%12\relax}%
\providecommand \@@startlink[1]{}%
\providecommand \@@endlink[0]{}%
\providecommand \url  [0]{\begingroup\@sanitize@url \@url }%
\providecommand \@url [1]{\endgroup\@href {#1}{\urlprefix }}%
\providecommand \urlprefix  [0]{URL }%
\providecommand \Eprint [0]{\href }%
\providecommand \doibase [0]{http://dx.doi.org/}%
\providecommand \selectlanguage [0]{\@gobble}%
\providecommand \bibinfo  [0]{\@secondoftwo}%
\providecommand \bibfield  [0]{\@secondoftwo}%
\providecommand \translation [1]{[#1]}%
\providecommand \BibitemOpen [0]{}%
\providecommand \bibitemStop [0]{}%
\providecommand \bibitemNoStop [0]{.\EOS\space}%
\providecommand \EOS [0]{\spacefactor3000\relax}%
\providecommand \BibitemShut  [1]{\csname bibitem#1\endcsname}%
\let\auto@bib@innerbib\@empty
%</preamble>
\bibitem [{\citenamefont {Pincus}(1991)}]{pincus_mm1991}%
  \BibitemOpen
  \bibfield  {author} {\bibinfo {author} {\bibfnamefont {Philip}\ \bibnamefont
  {Pincus}},\ }\bibfield  {title} {\enquote {\bibinfo {title} {Colloid
  stabilization with grafted polyelectrolytes},}\ }\href {\doibase
  10.1021/ma00010a043} {\bibfield  {journal} {\bibinfo  {journal}
  {Macromolecules}\ }\textbf {\bibinfo {volume} {24}},\ \bibinfo {pages} {2912}
  (\bibinfo {year} {1991})}\BibitemShut {NoStop}%
\bibitem [{\citenamefont {Russel}\ \emph {et~al.}(1989)\citenamefont {Russel},
  \citenamefont {Saville},\ and\ \citenamefont {Showalter}}]{saville_book}%
  \BibitemOpen
  \bibfield  {author} {\bibinfo {author} {\bibfnamefont {W.~B.}\ \bibnamefont
  {Russel}}, \bibinfo {author} {\bibfnamefont {D.~A.}\ \bibnamefont {Saville}},
  \ and\ \bibinfo {author} {\bibfnamefont {W.~R.}\ \bibnamefont {Showalter}},\
  }\href@noop {} {\emph {\bibinfo {title} {Colloidal Dispersions}}}\ (\bibinfo
  {publisher} {Cambridge University Press},\ \bibinfo {address} {Cambridge,
  England},\ \bibinfo {year} {1989})\BibitemShut {NoStop}%
\bibitem [{\citenamefont {Derjaguin}\ and\ \citenamefont
  {Landau}(1941)}]{dlvo1}%
  \BibitemOpen
  \bibfield  {author} {\bibinfo {author} {\bibfnamefont {B.~V.}\ \bibnamefont
  {Derjaguin}}\ and\ \bibinfo {author} {\bibfnamefont {L.~D.}\ \bibnamefont
  {Landau}},\ }\bibfield  {title} {\enquote {\bibinfo {title} {Theory of the
  stability of strongly charged lyophobic sols and the adhesion of strongly
  charged particles in solutions of electrolytes},}\ }\href@noop {} {\bibfield
  {journal} {\bibinfo  {journal} {Acta Physicochim URSS}\ }\textbf {\bibinfo
  {volume} {14}},\ \bibinfo {pages} {633} (\bibinfo {year} {1941})}\BibitemShut
  {NoStop}%
\bibitem [{\citenamefont {Verwey}\ and\ \citenamefont
  {Overbeek}(1948)}]{dlvo2}%
  \BibitemOpen
  \bibfield  {author} {\bibinfo {author} {\bibfnamefont {E.~J.~W.}\
  \bibnamefont {Verwey}}\ and\ \bibinfo {author} {\bibfnamefont {J.~Th.~G.}\
  \bibnamefont {Overbeek}},\ }\href@noop {} {\emph {\bibinfo {title} {Theory of
  the Stability of Lyophobic Colloids}}}\ (\bibinfo  {publisher} {Elsevier},\
  \bibinfo {address} {Amsterdam},\ \bibinfo {year} {1948})\BibitemShut
  {NoStop}%
\bibitem [{\citenamefont {Yethiraj}\ and\ \citenamefont {van
  Blaaderen}(2003)}]{blaaderen_nature2003}%
  \BibitemOpen
  \bibfield  {author} {\bibinfo {author} {\bibfnamefont {Anand}\ \bibnamefont
  {Yethiraj}}\ and\ \bibinfo {author} {\bibfnamefont {Alfons}\ \bibnamefont
  {van Blaaderen}},\ }\bibfield  {title} {\enquote {\bibinfo {title} {A
  colloidal model system with an interaction tunable from hard sphere to soft
  and dipolar},}\ }\href {\doibase 10.1038/nature01328} {\bibfield  {journal}
  {\bibinfo  {journal} {Nature (London)}\ }\textbf {\bibinfo {volume} {421}},\
  \bibinfo {pages} {513} (\bibinfo {year} {2003})}\BibitemShut {NoStop}%
\bibitem [{\citenamefont {Leunissen}\ \emph {et~al.}(2005)\citenamefont
  {Leunissen}, \citenamefont {Christova}, \citenamefont {Hynninen},
  \citenamefont {Royall}, \citenamefont {Campbell}, \citenamefont {Imhof},
  \citenamefont {Dijkstra}, \citenamefont {van Roij},\ and\ \citenamefont {van
  Blaaderen}}]{blaaderen_nature2005}%
  \BibitemOpen
  \bibfield  {author} {\bibinfo {author} {\bibfnamefont {Mirjam~E.}\
  \bibnamefont {Leunissen}}, \bibinfo {author} {\bibfnamefont {Christina~G.}\
  \bibnamefont {Christova}}, \bibinfo {author} {\bibfnamefont {Antti-Pekka}\
  \bibnamefont {Hynninen}}, \bibinfo {author} {\bibfnamefont {C.~Patrick}\
  \bibnamefont {Royall}}, \bibinfo {author} {\bibfnamefont {Andrew~I.}\
  \bibnamefont {Campbell}}, \bibinfo {author} {\bibfnamefont {Arnout}\
  \bibnamefont {Imhof}}, \bibinfo {author} {\bibfnamefont {Marjolein}\
  \bibnamefont {Dijkstra}}, \bibinfo {author} {\bibfnamefont {Ren\'{e}}\
  \bibnamefont {van Roij}}, \ and\ \bibinfo {author} {\bibfnamefont {Alfons}\
  \bibnamefont {van Blaaderen}},\ }\bibfield  {title} {\enquote {\bibinfo
  {title} {Ionic colloidal crystals of oppositely charged particles},}\ }\href
  {\doibase 10.1038/nature03946} {\bibfield  {journal} {\bibinfo  {journal}
  {Nature (London)}\ }\textbf {\bibinfo {volume} {437}},\ \bibinfo {pages}
  {235} (\bibinfo {year} {2005})}\BibitemShut {NoStop}%
\bibitem [{\citenamefont {Sadakane}\ \emph {et~al.}(2009)\citenamefont
  {Sadakane}, \citenamefont {Onuki}, \citenamefont {Nishida}, \citenamefont
  {Koizumi},\ and\ \citenamefont {Seto}}]{sadakane2009}%
  \BibitemOpen
  \bibfield  {author} {\bibinfo {author} {\bibfnamefont {Koichiro}\
  \bibnamefont {Sadakane}}, \bibinfo {author} {\bibfnamefont {Akira}\
  \bibnamefont {Onuki}}, \bibinfo {author} {\bibfnamefont {Koji}\ \bibnamefont
  {Nishida}}, \bibinfo {author} {\bibfnamefont {Satoshi}\ \bibnamefont
  {Koizumi}}, \ and\ \bibinfo {author} {\bibfnamefont {Hideki}\ \bibnamefont
  {Seto}},\ }\bibfield  {title} {\enquote {\bibinfo {title} {Multilamellar
  structures induced by hydrophilic and hydrophobic ions added to a binary
  mixture of ${\mathbf{d}}_{2}\mathbf{O}$ and 3-methylpyridine},}\ }\href
  {\doibase 10.1103/PhysRevLett.103.167803} {\bibfield  {journal} {\bibinfo
  {journal} {Phys. Rev. Lett.}\ }\textbf {\bibinfo {volume} {103}},\ \bibinfo
  {pages} {167803} (\bibinfo {year} {2009})}\BibitemShut {NoStop}%
\bibitem [{\citenamefont {Sadakane}\ \emph {et~al.}(2013)\citenamefont
  {Sadakane}, \citenamefont {Nagao}, \citenamefont {Endo},\ and\ \citenamefont
  {Seto}}]{sadakane2013}%
  \BibitemOpen
  \bibfield  {author} {\bibinfo {author} {\bibfnamefont {Koichiro}\
  \bibnamefont {Sadakane}}, \bibinfo {author} {\bibfnamefont {Michihiro}\
  \bibnamefont {Nagao}}, \bibinfo {author} {\bibfnamefont {Hitoshi}\
  \bibnamefont {Endo}}, \ and\ \bibinfo {author} {\bibfnamefont {Hideki}\
  \bibnamefont {Seto}},\ }\bibfield  {title} {\enquote {\bibinfo {title}
  {Membrane formation by preferential solvation of ions in mixture of water,
  3-methylpyridine, and sodium tetraphenylborate},}\ }\href {\doibase
  http://dx.doi.org/10.1063/1.4838795} {\bibfield  {journal} {\bibinfo
  {journal} {J. Chem. Phys.}\ }\textbf {\bibinfo {volume} {139}},\ \bibinfo
  {eid} {234905} (\bibinfo {year} {2013})}\BibitemShut {NoStop}%
\bibitem [{\citenamefont {Leunissen}\ \emph
  {et~al.}(2007{\natexlab{a}})\citenamefont {Leunissen}, \citenamefont {van
  Blaaderen}, \citenamefont {Hollingsworth}, \citenamefont {Sullivan},\ and\
  \citenamefont {Chaikin}}]{chaikin_pnas2007}%
  \BibitemOpen
  \bibfield  {author} {\bibinfo {author} {\bibfnamefont {Mirjam~E.}\
  \bibnamefont {Leunissen}}, \bibinfo {author} {\bibfnamefont {Alfons}\
  \bibnamefont {van Blaaderen}}, \bibinfo {author} {\bibfnamefont {Andrew~D.}\
  \bibnamefont {Hollingsworth}}, \bibinfo {author} {\bibfnamefont {Matthew~T.}\
  \bibnamefont {Sullivan}}, \ and\ \bibinfo {author} {\bibfnamefont {Paul~M.}\
  \bibnamefont {Chaikin}},\ }\bibfield  {title} {\enquote {\bibinfo {title}
  {Electrostatics at the oil-water interface, stability, and order in emulsions
  and colloids},}\ }\href@noop {} {\bibfield  {journal} {\bibinfo  {journal}
  {Proc. Natl. Acad. Sci. U.S.A.}\ }\textbf {\bibinfo {volume} {104}},\
  \bibinfo {pages} {2585} (\bibinfo {year} {2007}{\natexlab{a}})}\BibitemShut
  {NoStop}%
\bibitem [{\citenamefont {Leunissen}\ \emph
  {et~al.}(2007{\natexlab{b}})\citenamefont {Leunissen}, \citenamefont
  {Zwanikken}, \citenamefont {van Roij}, \citenamefont {Chaikin},\ and\
  \citenamefont {van Blaaderen}}]{van_roij_pccp2007}%
  \BibitemOpen
  \bibfield  {author} {\bibinfo {author} {\bibfnamefont {Mirjam~E.}\
  \bibnamefont {Leunissen}}, \bibinfo {author} {\bibfnamefont {Jos}\
  \bibnamefont {Zwanikken}}, \bibinfo {author} {\bibfnamefont {Ren\'{e}}\
  \bibnamefont {van Roij}}, \bibinfo {author} {\bibfnamefont {Paul~M.}\
  \bibnamefont {Chaikin}}, \ and\ \bibinfo {author} {\bibfnamefont {Alfons}\
  \bibnamefont {van Blaaderen}},\ }\bibfield  {title} {\enquote {\bibinfo
  {title} {Ion partitioning at the oil-water interface as a source of tunable
  electrostatic effects in emulsions with colloids},}\ }\href@noop {}
  {\bibfield  {journal} {\bibinfo  {journal} {Phys. Chem. Chem. Phys.}\
  }\textbf {\bibinfo {volume} {9}},\ \bibinfo {pages} {6405} (\bibinfo {year}
  {2007}{\natexlab{b}})}\BibitemShut {NoStop}%
\bibitem [{\citenamefont {Zwanikken}\ and\ \citenamefont {van
  Roij}(2007)}]{van_roij_prl2007}%
  \BibitemOpen
  \bibfield  {author} {\bibinfo {author} {\bibfnamefont {Jos}\ \bibnamefont
  {Zwanikken}}\ and\ \bibinfo {author} {\bibfnamefont {Ren\'{e}}\ \bibnamefont
  {van Roij}},\ }\bibfield  {title} {\enquote {\bibinfo {title} {Charged
  colloidal particles and small mobile ions near the oil-water interface:
  Destruction of colloidal double layer and ionic charge separation},}\
  }\href@noop {} {\bibfield  {journal} {\bibinfo  {journal} {Phys. Rev. Lett.}\
  }\textbf {\bibinfo {volume} {99}},\ \bibinfo {pages} {178301} (\bibinfo
  {year} {2007})}\BibitemShut {NoStop}%
\bibitem [{\citenamefont {Samin}\ and\ \citenamefont {Tsori}(2013)}]{sesc_jcp}%
  \BibitemOpen
  \bibfield  {author} {\bibinfo {author} {\bibfnamefont {Sela}\ \bibnamefont
  {Samin}}\ and\ \bibinfo {author} {\bibfnamefont {Yoav}\ \bibnamefont
  {Tsori}},\ }\bibfield  {title} {\enquote {\bibinfo {title} {Stabilization of
  charged and neutral colloids in salty mixtures},}\ }\href@noop {} {\bibfield
  {journal} {\bibinfo  {journal} {J. Chem. Phys.}\ }\textbf {\bibinfo {volume}
  {139}},\ \bibinfo {pages} {244905} (\bibinfo {year} {2013})}\BibitemShut
  {NoStop}%
\bibitem [{\citenamefont {Marcus}(1985)}]{marcus_book1}%
  \BibitemOpen
  \bibfield  {author} {\bibinfo {author} {\bibfnamefont {Yizhak}\ \bibnamefont
  {Marcus}},\ }\href@noop {} {\emph {\bibinfo {title} {Ion Solvation}}}\
  (\bibinfo  {publisher} {Wiley},\ \bibinfo {address} {New York},\ \bibinfo
  {year} {1985})\BibitemShut {NoStop}%
\bibitem [{\citenamefont {Grattoni}\ \emph {et~al.}(1993)\citenamefont
  {Grattoni}, \citenamefont {Dawe}, \citenamefont {Seah},\ and\ \citenamefont
  {Gray}}]{grat93}%
  \BibitemOpen
  \bibfield  {author} {\bibinfo {author} {\bibfnamefont {Carlos~A.}\
  \bibnamefont {Grattoni}}, \bibinfo {author} {\bibfnamefont {Richard~A.}\
  \bibnamefont {Dawe}}, \bibinfo {author} {\bibfnamefont {C.~Yen}\ \bibnamefont
  {Seah}}, \ and\ \bibinfo {author} {\bibfnamefont {Jane~D.}\ \bibnamefont
  {Gray}},\ }\bibfield  {title} {\enquote {\bibinfo {title} {Lower critical
  solution coexistence curve and physical properties (density, viscosity,
  surface tension, and interfacial tension) of 2,6-lutidine + water},}\ }\href
  {\doibase 10.1021/je00012a008} {\bibfield  {journal} {\bibinfo  {journal} {J.
  Chem. Eng. Data}\ }\textbf {\bibinfo {volume} {38}},\ \bibinfo {pages} {516}
  (\bibinfo {year} {1993})}\BibitemShut {NoStop}%
\bibitem [{\citenamefont {Renard}\ and\ \citenamefont
  {Oberg}(1965)}]{renard1965}%
  \BibitemOpen
  \bibfield  {author} {\bibinfo {author} {\bibfnamefont {J.~A.}\ \bibnamefont
  {Renard}}\ and\ \bibinfo {author} {\bibfnamefont {A.~G.}\ \bibnamefont
  {Oberg}},\ }\bibfield  {title} {\enquote {\bibinfo {title} {Ternary systems:
  Water-acetonitrile-salts.}}\ }\href {\doibase 10.1021/je60025a025} {\bibfield
   {journal} {\bibinfo  {journal} {J. Chem. Eng. Data}\ }\textbf {\bibinfo
  {volume} {10}},\ \bibinfo {pages} {152} (\bibinfo {year} {1965})}\BibitemShut
  {NoStop}%
\bibitem [{\citenamefont {Safran}(1994)}]{safran_book}%
  \BibitemOpen
  \bibfield  {author} {\bibinfo {author} {\bibfnamefont {Samuel}\ \bibnamefont
  {Safran}},\ }\href@noop {} {\emph {\bibinfo {title} {Statistical
  Thermodynamics of Surfaces, Interfaces, and Membranes}}}\ (\bibinfo
  {publisher} {Westview Press},\ \bibinfo {address} {New York},\ \bibinfo
  {year} {1994})\BibitemShut {NoStop}%
\bibitem [{\citenamefont {Hunter}(2001)}]{hunter2001}%
  \BibitemOpen
  \bibfield  {author} {\bibinfo {author} {\bibfnamefont {Robert~J.}\
  \bibnamefont {Hunter}},\ }\href@noop {} {\emph {\bibinfo {title} {Foundations
  of Colloid Science}}},\ \bibinfo {edition} {2nd}\ ed.\ (\bibinfo  {publisher}
  {Oxford University},\ \bibinfo {address} {New York},\ \bibinfo {year}
  {2001})\BibitemShut {NoStop}%
\bibitem [{\citenamefont {Okamoto}\ and\ \citenamefont
  {Onuki}(2011)}]{onuki2011}%
  \BibitemOpen
  \bibfield  {author} {\bibinfo {author} {\bibfnamefont {Ryuichi}\ \bibnamefont
  {Okamoto}}\ and\ \bibinfo {author} {\bibfnamefont {Akira}\ \bibnamefont
  {Onuki}},\ }\bibfield  {title} {\enquote {\bibinfo {title} {Charged colloids
  in an aqueous mixture with a salt},}\ }\href@noop {} {\bibfield  {journal}
  {\bibinfo  {journal} {Phys. Rev. E}\ }\textbf {\bibinfo {volume} {84}},\
  \bibinfo {pages} {051401} (\bibinfo {year} {2011})}\BibitemShut {NoStop}%
\bibitem [{\citenamefont {Levin}(2009)}]{levin_prl2009a}%
  \BibitemOpen
  \bibfield  {author} {\bibinfo {author} {\bibfnamefont {Yan}\ \bibnamefont
  {Levin}},\ }\bibfield  {title} {\enquote {\bibinfo {title} {Polarizable ions
  at interfaces},}\ }\href {\doibase 10.1103/PhysRevLett.102.147803} {\bibfield
   {journal} {\bibinfo  {journal} {Phys. Rev. Lett.}\ }\textbf {\bibinfo
  {volume} {102}},\ \bibinfo {pages} {147803} (\bibinfo {year}
  {2009})}\BibitemShut {NoStop}%
\bibitem [{\citenamefont {Levin}\ \emph {et~al.}(2009)\citenamefont {Levin},
  \citenamefont {dos Santos},\ and\ \citenamefont {Diehl}}]{levin_prl2009b}%
  \BibitemOpen
  \bibfield  {author} {\bibinfo {author} {\bibfnamefont {Yan}\ \bibnamefont
  {Levin}}, \bibinfo {author} {\bibfnamefont {Alexandre~P.}\ \bibnamefont {dos
  Santos}}, \ and\ \bibinfo {author} {\bibfnamefont {Alexandre}\ \bibnamefont
  {Diehl}},\ }\bibfield  {title} {\enquote {\bibinfo {title} {Ions at the
  air-water interface: An end to a hundred-year-old mystery?}}\ }\href
  {\doibase 10.1103/PhysRevLett.103.257802} {\bibfield  {journal} {\bibinfo
  {journal} {Phys. Rev. Lett.}\ }\textbf {\bibinfo {volume} {103}},\ \bibinfo
  {pages} {257802} (\bibinfo {year} {2009})}\BibitemShut {NoStop}%
\bibitem [{\citenamefont {Nellen}\ \emph {et~al.}(2011)\citenamefont {Nellen},
  \citenamefont {Dietrich}, \citenamefont {Helden}, \citenamefont {Chodankar},
  \citenamefont {Nyg\aa{}rd}, \citenamefont {van~der Veen},\ and\ \citenamefont
  {Bechinger}}]{nellen2011}%
  \BibitemOpen
  \bibfield  {author} {\bibinfo {author} {\bibfnamefont {Ursula}\ \bibnamefont
  {Nellen}}, \bibinfo {author} {\bibfnamefont {Julian}\ \bibnamefont
  {Dietrich}}, \bibinfo {author} {\bibfnamefont {Laurent}\ \bibnamefont
  {Helden}}, \bibinfo {author} {\bibfnamefont {Shirish}\ \bibnamefont
  {Chodankar}}, \bibinfo {author} {\bibfnamefont {Kim}\ \bibnamefont
  {Nyg\aa{}rd}}, \bibinfo {author} {\bibfnamefont {J.~Friso}\ \bibnamefont
  {van~der Veen}}, \ and\ \bibinfo {author} {\bibfnamefont {Clemens}\
  \bibnamefont {Bechinger}},\ }\bibfield  {title} {\enquote {\bibinfo {title}
  {Salt-induced changes of colloidal interactions in critical mixtures},}\
  }\href@noop {} {\bibfield  {journal} {\bibinfo  {journal} {Soft Matter}\
  }\textbf {\bibinfo {volume} {7}},\ \bibinfo {pages} {5360} (\bibinfo {year}
  {2011})}\BibitemShut {NoStop}%
\bibitem [{\citenamefont {Netz}(1996)}]{netz_prl1996}%
  \BibitemOpen
  \bibfield  {author} {\bibinfo {author} {\bibfnamefont {Roland~R.}\
  \bibnamefont {Netz}},\ }\bibfield  {title} {\enquote {\bibinfo {title}
  {Colloidal flocculation in near-critical binary mixtures},}\ }\href {\doibase
  10.1103/PhysRevLett.76.3646} {\bibfield  {journal} {\bibinfo  {journal}
  {Phys. Rev. Lett.}\ }\textbf {\bibinfo {volume} {76}},\ \bibinfo {pages}
  {3646--3649} (\bibinfo {year} {1996})}\BibitemShut {NoStop}%
\bibitem [{\citenamefont {Samin}\ and\ \citenamefont
  {Tsori}(2011)}]{efips_epl}%
  \BibitemOpen
  \bibfield  {author} {\bibinfo {author} {\bibfnamefont {S.}~\bibnamefont
  {Samin}}\ and\ \bibinfo {author} {\bibfnamefont {Y.}~\bibnamefont {Tsori}},\
  }\bibfield  {title} {\enquote {\bibinfo {title} {Attraction between
  like-charge surfaces in polar mixtures},}\ }\href@noop {} {\bibfield
  {journal} {\bibinfo  {journal} {Europhys. Lett.}\ }\textbf {\bibinfo {volume}
  {95}},\ \bibinfo {pages} {36002} (\bibinfo {year} {2011})}\BibitemShut
  {NoStop}%
\bibitem [{\citenamefont {Bier}\ \emph {et~al.}(2011)\citenamefont {Bier},
  \citenamefont {Gambassi}, \citenamefont {Oettel},\ and\ \citenamefont
  {Dietrich}}]{bier2011}%
  \BibitemOpen
  \bibfield  {author} {\bibinfo {author} {\bibfnamefont {M.}~\bibnamefont
  {Bier}}, \bibinfo {author} {\bibfnamefont {A.}~\bibnamefont {Gambassi}},
  \bibinfo {author} {\bibfnamefont {M.}~\bibnamefont {Oettel}}, \ and\ \bibinfo
  {author} {\bibfnamefont {S.}~\bibnamefont {Dietrich}},\ }\bibfield  {title}
  {\enquote {\bibinfo {title} {Electrostatic interactions in critical
  solvents},}\ }\href@noop {} {\bibfield  {journal} {\bibinfo  {journal} {EPL}\
  }\textbf {\bibinfo {volume} {95}},\ \bibinfo {pages} {60001} (\bibinfo {year}
  {2011})}\BibitemShut {NoStop}%
\bibitem [{\citenamefont {Inerowicz}\ \emph {et~al.}(1994)\citenamefont
  {Inerowicz}, \citenamefont {Li},\ and\ \citenamefont
  {Persson}}]{persson1990}%
  \BibitemOpen
  \bibfield  {author} {\bibinfo {author} {\bibfnamefont {Halina~D.}\
  \bibnamefont {Inerowicz}}, \bibinfo {author} {\bibfnamefont {Wei}\
  \bibnamefont {Li}}, \ and\ \bibinfo {author} {\bibfnamefont {Ingmar}\
  \bibnamefont {Persson}},\ }\bibfield  {title} {\enquote {\bibinfo {title}
  {Determination of the transfer thermodynamic functions for some monovalent
  ions from water to \textit{N{,}N}-dimethylthioformamide{,} and for some
  anions from water to methanol{,} dimethyl sulfoxide{,} acetonitrile and
  pyridine{,} and standard electrode potentials of some
  \uppercase{m}+/\uppercase{m}(s) couples in
  \textit{N{,}N}-dimethylthioformamide},}\ }\href {\doibase
  10.1039/FT9949002223} {\bibfield  {journal} {\bibinfo  {journal} {J. Chem.
  Soc. Faraday Trans.}\ }\textbf {\bibinfo {volume} {90}},\ \bibinfo {pages}
  {2223} (\bibinfo {year} {1994})}\BibitemShut {NoStop}%
\bibitem [{com()}]{comment}%
  \BibitemOpen
  \href@noop {} {}\bibinfo {howpublished} {2,6-lutidine is a structural analog
  of pyridine. Hence, Gibbs transfer energies are estimated from the data for a
  water-pyridine mixture in Ref. \cite{persson1990}.}\BibitemShut {Stop}%
\bibitem [{\citenamefont {Landau}\ \emph {et~al.}(1984)\citenamefont {Landau},
  \citenamefont {Lifshitz},\ and\ \citenamefont {Pitaevskii}}]{landau2}%
  \BibitemOpen
  \bibfield  {author} {\bibinfo {author} {\bibfnamefont {L.~D.}\ \bibnamefont
  {Landau}}, \bibinfo {author} {\bibfnamefont {E.~M}\ \bibnamefont {Lifshitz}},
  \ and\ \bibinfo {author} {\bibfnamefont {L.~P.}\ \bibnamefont {Pitaevskii}},\
  }\href@noop {} {\emph {\bibinfo {title} {Electrodynamics of Continuous
  Media}}},\ \bibinfo {edition} {2nd}\ ed.\ (\bibinfo  {publisher}
  {Butterworth-Heinemann},\ \bibinfo {address} {Amsterdam},\ \bibinfo {year}
  {1984})\BibitemShut {NoStop}%
\bibitem [{\citenamefont {Ben-Yaakov}\ \emph {et~al.}(2009)\citenamefont
  {Ben-Yaakov}, \citenamefont {David}, \citenamefont {Harries},\ and\
  \citenamefont {Podgornik}}]{andelman_jpcb2009}%
  \BibitemOpen
  \bibfield  {author} {\bibinfo {author} {\bibfnamefont {Dan}\ \bibnamefont
  {Ben-Yaakov}}, \bibinfo {author} {\bibnamefont {David}}, \bibinfo {author}
  {\bibfnamefont {Andelman~Daniel}\ \bibnamefont {Harries}}, \ and\ \bibinfo
  {author} {\bibfnamefont {Rudi}\ \bibnamefont {Podgornik}},\ }\bibfield
  {title} {\enquote {\bibinfo {title} {Ions in mixed dielectric solvents:
  Density profiles and osmotic pressure between charged interfaces},}\
  }\href@noop {} {\bibfield  {journal} {\bibinfo  {journal} {J. Phys. Chem. B}\
  }\textbf {\bibinfo {volume} {113}},\ \bibinfo {pages} {6001} (\bibinfo {year}
  {2009})}\BibitemShut {NoStop}%
\bibitem [{\citenamefont {Law}\ \emph {et~al.}(1998)\citenamefont {Law},
  \citenamefont {Petit},\ and\ \citenamefont {Beysens}}]{Beysens1998}%
  \BibitemOpen
  \bibfield  {author} {\bibinfo {author} {\bibfnamefont {B.~M.}\ \bibnamefont
  {Law}}, \bibinfo {author} {\bibfnamefont {J.-M.}\ \bibnamefont {Petit}}, \
  and\ \bibinfo {author} {\bibfnamefont {D.}~\bibnamefont {Beysens}},\
  }\bibfield  {title} {\enquote {\bibinfo {title} {Adsorption-induced
  reversible colloidal aggregation},}\ }\href@noop {} {\bibfield  {journal}
  {\bibinfo  {journal} {Phys. Rev. E}\ }\textbf {\bibinfo {volume} {57}},\
  \bibinfo {pages} {5782} (\bibinfo {year} {1998})}\BibitemShut {NoStop}%
\bibitem [{\citenamefont {Parsegian}(2005)}]{parsegian_book}%
  \BibitemOpen
  \bibfield  {author} {\bibinfo {author} {\bibfnamefont {V.~A.}\ \bibnamefont
  {Parsegian}},\ }\href@noop {} {\emph {\bibinfo {title} {Van der Waals Forces:
  A Handbook for Biologists, Chemists, Engineers, and Physicists}}}\ (\bibinfo
  {publisher} {Cambridge University Press},\ \bibinfo {address} {Cambridge,
  England},\ \bibinfo {year} {2005})\BibitemShut {NoStop}%
\bibitem [{\citenamefont {Bier}\ \emph {et~al.}(2012)\citenamefont {Bier},
  \citenamefont {Gambassi},\ and\ \citenamefont {Dietrich}}]{bier2012}%
  \BibitemOpen
  \bibfield  {author} {\bibinfo {author} {\bibfnamefont {Markus}\ \bibnamefont
  {Bier}}, \bibinfo {author} {\bibfnamefont {Andrea}\ \bibnamefont {Gambassi}},
  \ and\ \bibinfo {author} {\bibfnamefont {Siegfried}\ \bibnamefont
  {Dietrich}},\ }\bibfield  {title} {\enquote {\bibinfo {title} {Local theory
  for ions in binary liquid mixtures},}\ }\href {\doibase 10.1063/1.4733973}
  {\bibfield  {journal} {\bibinfo  {journal} {J. Chem. Phys.}\ }\textbf
  {\bibinfo {volume} {137}},\ \bibinfo {eid} {034504} (\bibinfo {year}
  {2012})}\BibitemShut {NoStop}%
\bibitem [{\citenamefont {Bai}\ \emph {et~al.}(2007)\citenamefont {Bai},
  \citenamefont {Huang}, \citenamefont {Yang},\ and\ \citenamefont
  {Huang}}]{bai2007}%
  \BibitemOpen
  \bibfield  {author} {\bibinfo {author} {\bibfnamefont {Feng}\ \bibnamefont
  {Bai}}, \bibinfo {author} {\bibfnamefont {Bo}~\bibnamefont {Huang}}, \bibinfo
  {author} {\bibfnamefont {Xinlin}\ \bibnamefont {Yang}}, \ and\ \bibinfo
  {author} {\bibfnamefont {Wenqiang}\ \bibnamefont {Huang}},\ }\bibfield
  {title} {\enquote {\bibinfo {title} {Synthesis of monodisperse porous
  poly(divinylbenzene) microspheres by distillation-precipitation
  polymerization},}\ }\href {\doibase 10.1016/j.polymer.2007.05.006} {\bibfield
   {journal} {\bibinfo  {journal} {Polymer}\ }\textbf {\bibinfo {volume}
  {48}},\ \bibinfo {pages} {3641} (\bibinfo {year} {2007})}\BibitemShut
  {NoStop}%
\bibitem [{sup()}]{supplementary}%
  \BibitemOpen
  \href@noop {} {\enquote {\bibinfo {title} {See supplemental material at
  http://link.aps.org/ supplemental/10.1103/physrevapplied.2.024008 for
  information regarding experimental details and procedures and numerical
  calculations.}}\ }\BibitemShut {NoStop}%
\bibitem [{\citenamefont {Buzaglo}\ \emph {et~al.}(2013)\citenamefont
  {Buzaglo}, \citenamefont {Shtein}, \citenamefont {Kober}, \citenamefont
  {Lovrincic}, \citenamefont {Vilan},\ and\ \citenamefont {Regev}}]{regev2013}%
  \BibitemOpen
  \bibfield  {author} {\bibinfo {author} {\bibfnamefont {Matat}\ \bibnamefont
  {Buzaglo}}, \bibinfo {author} {\bibfnamefont {Michael}\ \bibnamefont
  {Shtein}}, \bibinfo {author} {\bibfnamefont {Sivan}\ \bibnamefont {Kober}},
  \bibinfo {author} {\bibfnamefont {Robert}\ \bibnamefont {Lovrincic}},
  \bibinfo {author} {\bibfnamefont {Ayelet}\ \bibnamefont {Vilan}}, \ and\
  \bibinfo {author} {\bibfnamefont {Oren}\ \bibnamefont {Regev}},\ }\bibfield
  {title} {\enquote {\bibinfo {title} {Critical parameters in exfoliating
  graphite into graphene},}\ }\href {\doibase 10.1039/C3CP43205J} {\bibfield
  {journal} {\bibinfo  {journal} {Phys. Chem. Chem. Phys.}\ }\textbf {\bibinfo
  {volume} {15}},\ \bibinfo {pages} {4428} (\bibinfo {year}
  {2013})}\BibitemShut {NoStop}%
\bibitem [{\citenamefont {Evans}\ \emph {et~al.}(1986)\citenamefont {Evans},
  \citenamefont {Marconi},\ and\ \citenamefont {Tarazona}}]{evans1986}%
  \BibitemOpen
  \bibfield  {author} {\bibinfo {author} {\bibfnamefont {R.}~\bibnamefont
  {Evans}}, \bibinfo {author} {\bibfnamefont {U.~Marini~Bettolo}\ \bibnamefont
  {Marconi}}, \ and\ \bibinfo {author} {\bibfnamefont {P.}~\bibnamefont
  {Tarazona}},\ }\bibfield  {title} {\enquote {\bibinfo {title} {Fluids in
  narrow pores: Adsorption, capillary condensation, and critical points},}\
  }\href {\doibase 10.1063/1.450352} {\bibfield  {journal} {\bibinfo  {journal}
  {J. Chem. Phys.}\ }\textbf {\bibinfo {volume} {84}},\ \bibinfo {pages} {2376}
  (\bibinfo {year} {1986})}\BibitemShut {NoStop}%
\end{thebibliography}


\end{document}